\documentclass[11pt,reqno]{amsart}

\usepackage[english]{babel}
\usepackage{graphicx}
\usepackage{textcomp}
\usepackage{amsmath}
\newtheorem{theorem}{ Theorem}[section]
\newtheorem{remark}{ Remark}[section]

\usepackage{url}
\usepackage{amsfonts,amssymb,amsmath,amsbsy,amsthm,amscd}

\begin{document}

\title [The stability of vortices in gas on the $l$-plane]
{The stability of vortices in gas on the $l$-plane: the influence of
centrifugal force}

%\title{The non-viscous Burgers equation in a noisy medium: a threshold effect for blow up behaviour}

\author[Rozanova,\,Turzynsky]{Olga Rozanova, Marko Turzynsky}

%\author[label1,label2]{}

\address{Mathematics and Mechanics Faculty, Moscow State University, Moscow
119991, Russia}

\email{rozanova@mech.math.msu.su}

%\subjclass {35R60}


%\date{\today}

\begin{abstract}
We show that a small correction due to centrifugal force usually
neglected in the  $l$-plane model of atmosphere drastically
influences on the stability of vortices. Namely, in the presence of
the Coriolis force only there exists a wide range of parameter
ensuring nonlinear stability of a vortex with uniform deformation.
Taking into account the centrifugal force results in a disappearance
of stable vortices in the above mentioned class of motions.

We also prove that for the heat ratio $\gamma=2$, corresponding to
the one-atomic gas, the system of  equations, describing the gas on
the $l$-plane with the correction due to centrifugal force can be
integrated in a special case.
\end{abstract}
%\pacs{}
\maketitle

\medskip

%\documentclass{svproc}
%
% RECOMMENDED %%%%%%%%%%%%%%%%%%%%%%%%%%%%%%%%%%%%%%%%%%%%%%%%%%%
%
%\usepackage[english,russian]{babel}
% to typeset URLs, URIs, and DOIs

%\institution{Moscow State University, Moscow, 119991, Russian
%Federation,} \email{rozanova@mech.math.msu.su}

%\maketitle              % typeset the title of the contribution

%
\section{Reduction of 3D primitive equations of the atmosphere dynamics to the $l$-plane model}

We consider the system of gas dynamics, widely used for a
description of the dynamics of atmosphere of a rotating planet, such
as the Earth. We do not dwell on the processes of heat and moisture
transfer and study the behavior of atmosphere in the frame of the
model of ideal polytropic inviscid gas. The model consists of the
following equations \cite{Vallis}
\begin{equation}\label{e1}
 \rho(\partial_t \bar U+ ( \bar U,\nabla) \bar U +2 \,\bar \omega\times \bar U +\bar \omega\times
  (\bar \omega\times\bar r)
 + g \bar e_3)=
-\nabla p,
\end{equation}
\begin{equation}\label{e2}
 \partial_t \rho + {\rm div}(\rho\bar U)=0,
 \end{equation}
 \begin{equation}\label{e3}
 \partial_t S +(\bar U,\nabla S)=0,
\end{equation}
for $\rho\ge 0$, $p\ge 0$, $\bar U=(U_1, U_2, U_3)$, $S$ (density,
pressure, velocity and entropy).
 The functions depend on time $t$
and on point $x\in {\mathbb R}^3,$ $\bar e_3=(0,0,1)$ is the
"upward" unit vector, $g$ is the acceleration due to gravity (points
in $-e_3$ direction to the center of planet), $\bar \omega$ is the
velocity of angular rotation, $\bar r$
is the radius-vector of point in the coordinate system, associated with the center of planet.   % $U_H=(U_1, U_2, 0)$.
The state equation is
\begin{equation}\label{e_state}
p=\rho^\gamma e^S,\qquad \gamma>1.
%\in (1, 2].
\end{equation}

System \eqref{e1} - \eqref{e3} describes
%relatively
%thin
 a layer of air over a spherical planet, therefore for the
processes of planetary scale one should use spherical coordinates.
Nevertheless, for the process of so called middle scale it is
possible to fix a point on the Earth surface and consider a
coordinate system with the origin in these point. Thus, axes $x_1$
and $x_2$ are directed in the tangent plane along parallel and
meridian, respectively, and the axis $x_3$ is directed out of the
center of planet.

Let us compute the term $\bar\omega\times (\bar\omega \times \bar
r)$ for the point lying on the plane $x_3=0$. Thus, the coordinates
of the radius-vector $\bar r$ are $(x_1, x_2, R)$, where $R$ is the
radius of planet. If we denote $\theta$ the latitude of the origin
of the rotating coordinate system and take into account that $\bar
\omega=(0, \omega \cos \theta, \omega \sin \theta)$, $\omega =|\bar
\omega|$,  then
\begin{equation}\nonumber
\bar \omega\times (\bar \omega \times \bar r)=R\omega^2 (0,\sin
\theta \,\cos \theta, -\cos^2\theta)- \omega^2 (x_1, x_2 \sin^2
\theta, x_2 \sin\theta \cos \theta).
\end{equation}
The first vector in this sum is constant, the second one depends on
the position of point on the plane $x_3=0$.

 Since $\bar\omega\times (\bar\omega \times
\bar r) = - \frac12\nabla  \omega^2 |\bar r|^2$, traditionally they
include this term in the geopotential $\Phi=g x_3-\frac12 \omega^2
|r|^2$ and turn the axis $(x_2, x_3)$ such that  the constant vector
$$R\omega^2 (0,\sin \theta \,\cos \theta, -\cos^2\theta)-(0,0,g)$$ in
the new coordinate system $(x_1, \tilde x_2, \tilde x_3)$ is
directed strictly vertically. In fact, we should turn the coordinate
system by the angle $\xi$ such that
$$
\cos \xi (-R \omega^2 \sin \theta \cos \theta)- \sin \xi
(-g+R\omega^2 \cos^2 \theta)=0,
$$
therefore
$$\tan \xi=\frac{R \omega^2 \sin \theta \cos \theta}{g-R\omega^2 \cos^2 \theta}.$$
We can see that $\xi \ll 1$, since $\frac{R\omega^2}{g}\ll 1$.

Thus, we change the spherical surface of planet to the geopotential
surface and regard the surfaces of constant $\Phi$ as being true
spheres. The horizontal component of apparent gravity is then
identically zero  (see the details in \cite{Vallis}, Sec.2.2.1).

If we are exactly at the origin, the term $\omega^2 (x_1, x_2 \sin^2
\theta, x_2 \sin\theta \cos \theta)$ is zero. However, in the
general case
$$
k=\cos \xi  \sin^2 \theta+ \sin \xi  \sin\theta\cos \theta\ne
0,\quad k\approx \sin^2 \theta,
$$
and this means that the centrifugal force still has a projection on
the new horizontal coordinates $(x_1, \tilde x_2)$ as $(- \omega^2
x_1, -k \omega^2 \tilde x_2)$. Of course, one can include these
terms in the gradient of some new pressure, however this would
change the state equation.

Let us repeat the standard procedure of deriving the $l$-plane model
for relatively thin atmosphere, ignoring vertical processes (e.g.
\cite{Vallis}, Sec.2.3.1). They define a plane tangent to the
surface of the earth at a fixed latitude $\theta$, and then use a
Cartesian coordinate system $(x_1, x_2)$ to describe motion on that
plane (we will write $x_2$ instead of our previous $\tilde x_2$).
For small excursions on the plane, they fix the projection of  $\bar
\omega$ in the direction of the local vertical, ignore the
components of $\bar \omega$  in the horizontal direction, set the
component of velocity $U_3=0$,  and make the hydrostatic
approximation, which means that the gravitational term is assumed to
be balanced by the pressure gradient term. Thus the 3D vectorial
equation \eqref{e1} results in the following 2D vectorial equation:
\begin{eqnarray}\label{(1)}
 \rho(\partial_t {\bf u} +  ({\bf u}\cdot \nabla ){\bf u} +
 {\mathcal L}{\bf u}- \delta\omega^2 \mathcal D {\bf x}) + \nabla p  = 0,
\end{eqnarray}
where ${\bf u}(t,x)=(U_1,U_2)$, $x\in{\mathbb R}^2,$ $\bf x$ is a
radius-vector of point,
 $\mathcal L = l L$,
$L=\left(\begin{array}{cr} 0 & -1 \\
1 & 0
\end{array}\right)$,  $l= 2 \omega \sin \theta  >0$ is the Coriolis
parameter, $\mathcal D=\left(\begin{array}{rr} 1 & 0 \\
0 & k
\end{array}\right)$, $k=\sin^2 \theta$. Equations \eqref{e2}, \eqref{e3} and
\eqref{e_state} are not  changed formally, however, the density,
pressure and entropy depend now only on $x\in{\mathbb R}^2$ and $t$.
In the isentropic case the equation \eqref{e3} is valid identically.
The value of $\delta$ is zero if we do not consider the centrifugal
term or one, otherwise.

The reduction to the 2D system for a thin layer of atmosphere can be
obtained more strictly by averaging over the height, see
\cite{Obukhov}. The case $\gamma=2$ corresponds to the  2D
one-atomic gas ($\gamma=2=1+\frac{1}{n}$, where $n$ is the dimension
of space). Also it arises in the shallow water equations.

\section{Motion with uniform deformation: steady states and their stability}

As is well known, the gas dynamics system has a special class of
solutions, characterized be a linear profile of velocity, that is
\begin{equation}\label{LP}
{\bf u}(t,{\bf x})=Q {\bf x}, \quad Q= \left(\begin{array}{cr} a(t) & b(t) \\
c(t)& d(t)\end{array}\right).
\end{equation}
 In \cite{RT1}  the system \eqref{(1)}, \eqref{e2}, \eqref{e3} was considered for
 $\delta=0$ and it was shown that the components of matrix
  $Q$ can be found as a part of solution of a certain
  quadratically-nonlinear matrix system of equations. For the case
  of arbitrary $\delta$ this system has the form
\begin{eqnarray}\label{(2.1)}
\dot R+ RQ + Q^T R+(\gamma-1){\bf tr}Q R =0,\\
\dot Q+Q^2+lLQ+2c_0 R-\delta \omega^2 \mathcal D=0,\label{(2.2)}
\end{eqnarray}
where the symmetric matrix  $R$ describes distribution of density
and pressure, $c_0=\rm const$.

\subsection{$\delta$=0}

System \eqref{(2.1)}, \eqref{(3)} has an one-parametric family of
nontrivial equilibria, depending on the parameter $b_*$:
\begin{eqnarray}\label{(3)}
a=d=0,\, b=-c=b_*,\, A=C=\frac{b_*(b_*-l)}{2c_0},\, B=0,
\end{eqnarray}
and a two-parametric family with parameters $a_*$ and $c_*$
\begin{eqnarray}\label{(3.1)}
a = a_*,\, c = c_*,\, b = -\frac{a_*^2}{c_*},\, d = -a_*,\, A =
\frac{ l c_*}{2 c_0} , \, B = -\frac{l a_*}{c_0},\, C =
\frac{la_*^2}{2 c_0 c_* },
\end{eqnarray}
for $c_*=0$  the latter equilibrium should be completed by
\begin{eqnarray}\label{(3.2)}
a = c=d= 0,\, b = b_*, \, A = B = 0,\, C = -\frac{1}{ 2 c_0} l b_*,
\end{eqnarray}
with a parameter $b_*$.

Equilibrium \eqref{(3)} corresponds to the case of axisymmetric
vortex. For $b_*\in (0,l)$ the motion is anticyclonic with a higher
pressure in the center of  vortex (anticyclonic high), for $b_*<0$
the motion is cyclonic with a lower pressure in the center of vortex
(cyclonic low), and at last, for $b_*>l$ the motion is anticyclonic
with a higher pressure in the center (anticyclonic low).

Equilibria \eqref{(3.1)}, \eqref{(3.2)} correspond to a shear flow
along a straight line, the ratio $\frac{a_*}{c_*}$ gives the
inclination of this line with respect to the coordinate axis.

Let us give a review of the results about stability of the
equilibria.
\begin{theorem}
Equilibrium \eqref{(3)} is unstable for any $\gamma>1$ if
\begin{equation}\label{c1}
\frac{b_*}{l}\in \mathbb R \backslash\Sigma, \,\mbox{where}\,
\Sigma=[K_-,K_+],\, K_\pm=\frac{1\pm\sqrt{2}}{2}.
\end{equation}
Moreover, for $\gamma>4$ the domain  $\Sigma$ can be replaced by
$\sigma=[k_-,k_+]$,
$k_\pm=\frac{1\pm\sqrt{{\gamma}/{(\gamma-2)}}}{2}$. Thus,
$\sigma=[k_-,k_+]\to [0,1]$ as $\gamma\to\infty$ and the domain of
instability enlarges.

\end{theorem}
\proof Eigenvalues of matrix of linearization in the point
\eqref{(3)} for the system \eqref{(2.1)}, \eqref{(2.2)}, written as
7 equations are the following:
$$\lambda_{1,2,3,4}=\pm\sqrt{
2}\,\sqrt{-l\left(b_*+\frac{l}{4}\right)\pm
\sqrt{\left(b_*+\frac{l}{2}\right)^2\left(\frac{l^2}{4}+b_*
l-(b_*)^2\right)}},$$ $$\lambda_{5,6}= \pm
\sqrt{-(2(2-\gamma)b_*(b_*-l)+l^2)},\qquad \lambda_{7}=0.$$

For $b_*< \frac{1-\sqrt{2}}{2}\,l$ and
$b_*>\frac{1+\sqrt{2}}{2}\,l>l$ we get  $\lambda_{3,4,5,6}=\pm
\alpha \pm i \beta, \,\alpha\ne 0, \beta \ne 0$, therefore there
exists an eigenvalue with a positive real part. Thus, the Lyapunov
theorem implies instability of the equilibrium.

For $\gamma\in(1,2]$ (this case was considered in \cite{RT1} ) the
value $-(2(2-\gamma)b_*(b_*-l)+l^2)<0$ , then
$\Re(\lambda_{1,2})=0$. For $\gamma>2$ we have
$-(2(2-\gamma)b_*(b_*-l)+l^2)>0$ (the eigenvalues $\lambda_{1,2}$
are real, and one of them is positive) for
\begin{equation}\label{c2}
\frac{b_*}{l}<k_- \quad\mbox{or}\quad \frac{b_*}{l}>k_+.
\end{equation}
It is easy to check that for $\gamma\in (2,4)$ condition \eqref{c1}
is more restrictive than \eqref{c2}, whereas for $\gamma>4$
condition \eqref{c2} is more restrictive. Therefore the proof is
over.$\square$

\bigskip

One can readily check that if $b_*\in
\big[\frac{1-\sqrt{2}}{2}\,l,\,\frac{1+\sqrt{2}}{2}\,l \big]$, then
all eigenvalues have zero real part. Thus, for this range of
parameters the linear analysis does not give an answer about
stability of the equilibrium. Nevertheless, the following theorem
holds.
\begin{theorem}\cite{RT1}
Equilibrium \eqref{(3)} is stable in the Lyapunov sense  for any
$\gamma>1$ if
\begin{equation}\nonumber\label{c3}
\frac{b_*}{l}\in (0,1).
\end{equation}
\end{theorem}

The proof is based on the explicit construction of the Lyapunov
function (see the details in\cite{RT1}).

 Very
recently the following theorem was proved.
\begin{theorem}\cite{T}
For $\gamma=2$ equilibrium \eqref{(3)} is stable in the Lyapunov
sense for all $$b_*\in (\tfrac{1-\sqrt 2}{2}\,l,\tfrac{1+\sqrt
2}{2}\,l),$$ and unstable otherwise.
\end{theorem}

 The proof is quite technical and based on the use of the
Lagrangian coordinates.

\bigskip
We do not dwell here on the study of stability of equilibria
\eqref{(3.1)}, \eqref{(3.2)}, it is not trivial.  The linear
analysis implies that the matrix of linearization has a couple of
eigenvalues  for $\pm \frac{1}{c_*}
\sqrt{c_*l((a_*^2+c_*^2)\gamma-c_*l)}$ for \eqref{(3.1)} and $\pm
\sqrt{-b_*l(\gamma+1)}$ for \eqref{(3.2)}, whereas the other
eigenvalues have the zero real part. Thus, one can easily find the
range of parameters guaranteing instability. The proof of
instability in the rest of cases is much more delicate.

\subsection{$\delta$=1, $s=1$}

To reduce computations and obtain analytical result, we dwell on a
particular case $s=1$. We have the following real equilibria:
\begin{equation}\label{d3}
a=d=0,\, b=-c=b_*,\, A=C= \frac{(\omega-b_*)^2}{2 c_0},\, B=0,
 \end{equation}
\begin{equation}\label{d3.1}
a = a_*,\,c = c_*, \, b = \frac{\omega^2-a_*^2}{c_*},\,  d = -a_*,\,
A =\frac{ c_*\omega}{c_0},\, B = -\frac{2a_*\omega}{c_0}, \,C
=\frac{ \omega(a_*^2-\omega^2)}{c_* c_0},\,
\end{equation}
for $c_*=0$ the later equilibrium should be completed by
\begin{equation}\label{d3.2}
a =\pm \omega, \,b = b_*,\, d = \mp \omega,
 A = 0,\, B = \mp\frac{2\omega^2}{c_0},\, C = -\frac{\omega b_*}{c_0}.\end{equation}
Thus, \eqref{d3} and \eqref{d3.1}, \eqref{d3.2} correspond to
\eqref{(3)} and \eqref{(3.1)}, \eqref{(3.2)} for $\delta=0$.

One can easily see that the "anticyclonic high" domain of parameters
disappears from \eqref{d3}. Thus, the situation is similar to the
case $\delta=0$ with $l=0$ (the plane does not rotate).

Equilibria \eqref{d3.1}, \eqref{d3.2} correspond to the saddle point
of pressure.

\begin{theorem}
Equilibrium \eqref{d3} is unstable for all values of parameter
$b_*$.
\end{theorem}

\proof The eigenvalues of matrix of linearization for the system
\eqref{(2.1)}, \eqref{(2.2)} at the point \eqref{d3} are the
following:
$$\lambda_1=0,\, \lambda_{2,3}=\pm \sqrt{2(\gamma-2)}(\omega-b_*),\,
\lambda_{4,5,6,7}=\pm \sqrt{-4b_*\omega\pm 2(b_*^2-\omega^2)i}$$
Therefore for $b_*\ne \omega$ there is an eigenvalue with a positive
real part. The only possibility to have the zero real part is
$b_*=\omega$. Let us prove that this case is also unstable. First of
all we notice that if $b_*=\omega$, than $A_*=C_*=0$. If we fix
$A=B=C=0$, we will get a subset of solutions of the full system
\eqref{(2.1)}, \eqref{(2.2)}, corresponding to $R=0$, that is
\begin{eqnarray}\nonumber
\dot Q+Q^2+2\omega LQ-\omega^2 \mathcal E=0,\label{(2.3)}
\end{eqnarray}
where $\mathcal E$ is the identity matrix. We are going to find a
small perturbation of equilibrium of this matrix Riccati equation
which results in a finite time blow up. For this raison we first
make a change $Q=\tilde Q-\omega L$, since $Q=-\omega L$ is the
steady state, corresponding to the equilibrium point. For $\tilde Q$
we get the matrix Bernoulli equation
\begin{eqnarray}
\dot {\tilde Q}=-\tilde Q^2+\omega \tilde Q L-\omega L \tilde
Q,\label{(2.4)}
\end{eqnarray}
where the equilibrium is shifted to the zero point. Equation
\eqref{(2.4)} can be reduced to the following linear equation for
$U=\tilde Q^{-1}$(see, e.g \cite{Egorov}):
\begin{eqnarray}
\dot U=\omega (UL-LU) +E.\label{(2.5)}
\end{eqnarray}
The initial condition $U(0)$ can be derived from $\tilde Q(0)$
provided $\det \tilde Q(0)\ne 0$. Equation \eqref{(2.5)} can be
explicitly solved. Let us choose  initial perturbations of the
equilibrium $\tilde Q=0$ as $\tilde Q_{11}=\tilde Q_{22}=0$, $\tilde
Q_{12}=\tilde Q_{21}=\varepsilon$. The components of $U(t)$ are
polynomials with respect to $t$ up to the second order (we skip the
standard computations), nevertheless, the components $\tilde
Q(t)=U^{-1} (t)$ contain $\det U(t)$ in the denominator. For our
specific choice of perturbation $\det U(t)=\left(
\frac{2\omega}{\varepsilon}+1\right)^2t-\frac{1}{\varepsilon^2}$ and
the components of  $\tilde Q(t)$ tend to infinity as $t\to
(2\omega+\varepsilon)^{-2}.$ Thus, the theorem is proved. $\square$

\bigskip

For  equilibria  \eqref{d3.1} and \eqref{d3.2} the spectra of
matrices of linearization contain three zeros and the quadriple
$$
\pm\frac{1}{c_*}\sqrt{c_*w(\gamma z_*\pm\sqrt{z_*^2\gamma^2+32
(\gamma-2) c_*w^2 })}, \qquad z_*=a_*^2+c_*^2-w^2
$$
(for \eqref{d3.1}) and
$$
\pm\sqrt{-b_*w\gamma \pm\omega\sqrt{b_*^2\gamma^2+32 (\gamma-2) w^2
})},
$$
(for \eqref{d3.2}). Thus, one can find conditions on the parameters
ensuring instability of  the equilibria. Let us notice that these
conditions strongly depend on the heat ratio $\gamma$. For example,
\eqref{d3.2} is always unstable for $\gamma>2$. The stability for
the linearly neutral cases is not studied, nevertheless, it is no
reason to expect stability for the saddle points of the pressure.

\subsection{$\delta$=1, $s=\sin\theta$}

In the general case the equilibria of system \eqref{(2.1)},
\eqref{(2.2)} can be found as roots of algebraic equations of a
higher order. We do not list them all, nevertheless let us mention
the following nontrivial family of equilibria depending only on one
parameter $c_*$:
\begin{eqnarray}\nonumber
a = 0,\, b = z , c = c_*,\, d = 0, \\\nonumber A=
 \frac{1}{2 c_0}(2 c_* w s-zc_*+w^2), \, B = 0,\,
 C = \frac{1}{2 c_0}(w^2 s^2-2 w s z-z c_*),
\end{eqnarray}
where $z=z(c_*,w)$ is a root of the quadratic equation
$$c_*z^2+(c_*
^2-w^2)z-c_*s^2w^2=0.$$ This equation always has two
real roots, therefore the properties of the equilibrium depend on
the choice of the root. One  equilibrium is a non-axisymmetric
vortex ($z=-c_*$), another one is a saddle point
($z=\frac{w^2}{c_*}$). If we set $s=1,$ we get \eqref{d3} and
\eqref{d3.1} for $a_*=0$.

For this case  analytical computations are very cumbersome,
nevertheless one can still find the eigenvalues of the matrix of
linearization and see that the spectrum always contains a quadriple
$\pm\alpha\pm  \beta i $ with a nonzero $\alpha$, therefore, the
equilibrium is unstable.

\section{Study of system \eqref{(2.1)}, \eqref{(2.2)} in the Lagrangian coordinates}

In the present section we use the Lagrangian coordinates to show
that for  $\gamma=2$ and $s=1$ the solution can be found in terms of
elliptic integrals.

Let $F=(F_{ik})_{i,k=1..2}$ be the matrix of transfer from the
Lagrangian coordinates  ${\bf w}$ to the Eulerian ones ${\bf x}$,
i.e. ${\bf x}(t,{\bf w})=F(t){\bf w}$. This property is called the
uniform deformation. Since ${\bf w}=F^{-1}{\bf x}$, then ${\bf
u}=\dot {\bf x}=\dot F{\bf w}=\dot F F^{-1}{\bf x}=Q{\bf x}$. Thus,
the condition of uniform deformation is equivalent to condition
\eqref{LP}.

System \eqref{(2.1)}, \eqref{(2.2)} can be written in the Lagrangian
coordinates as
\begin{eqnarray}\label{4}
\ddot F_{ik}+\frac{\partial U}{\partial F_{ik}}+\sum_{j}\mathcal
L_{ij}\dot F_{jk}-\delta \omega^{2}\mathcal D F_{ik}=0.
\end{eqnarray}
%see \cite{Ovsiannikov_book}.
Here $U$ is the internal energy of gas, $U=U_{0}(|\det
F|)^{-\gamma}\det F$, $U_{0}=\rm const>0$, $F$ is nondegenerate, the
derivatives are taken with respect to time.

%При заданных восьми начальных условиях $F_{ij}(0)$ и $\dot
%F_{ij}(0)$ по крайней мере локально по времени система (4) имеет
%единственное гладкое решение.

For $l=\delta=0$ system \eqref{4} was considered extensively in the
literature both in 3D and 2D cases. It models an expansion of a gas
ellipsoid (or elliptic cylinder in 2D case) into vacuum. The problem
originates from the paper of L.V.Ovsyannikov \cite{Ovsyannikov}
(1956), the state of art and respective references can be found in
\cite{BorisovKilinMamaev}. In \cite{Dyson}, \cite{AnisimovLysikov}
the first integrals of the system  \eqref{4} were found for
$l=\delta=0$. The integrals are the following:
$$\frac{1}{2}\sum_{i,k}(\dot F_{ik})^{2}+U=E,$$ $$\,F_{11}\dot
F_{21}+F_{12}\dot F_{22}-F_{21}\dot F_{11}-F_{22}\dot F_{12}=J,$$
$$\,F_{11}\dot F_{12}+F_{21}\dot F_{22}-F_{12}\dot F_{11}-F_{22}\dot
F_{21}=K.$$ Moreover, for the case of one-atomic gas (for $x\in
{\mathbb R}^2$ it corresponds to $\gamma=2$) in
\cite{AnisimovLysikov} it was shown the existence of a supplemental
 first integral
\begin{equation}\label{int_g2}
G=\sum_{i,k} F_{ik}^{2}=2Et^{2}+k_1t+k_0,
\end{equation}
 where $k_0$ and $k_1$ are
constants. It gives a possibility to find exact solution to system
\eqref{4} in the 2D case. In \cite{Bogoyavlensky} an oscillating
regime of \eqref{4} were established.

\subsection{First integrals}

%Введем обозначения: $\frac{1}{2}\sum_{i,k}(F'_{ik})^{2}+U=E;
%\,F_{11}F'_{21}+F_{12}F'_{22}-F_{21}F'_{11}-F_{22}F'_{12}=J;$
%$F_{11}F'_{12}+F_{21}F'_{22}-F_{12}F'_{11}-F_{22}F'_{21}=K.$
\begin{theorem} System \eqref{4}
has the integral of energy
\begin{equation}\label{5.0}
\tilde
E=E-\frac{1}{2}\omega^{2}(F_{11}^2+F_{12}^2+k(F_{21}^2+F_{22}^2)).
\end{equation}
For $s=1$ the system \eqref{4} has three first integrals:
\begin{equation}\label{5}
\tilde E=E-\frac{1}{2}\omega^{2}G,\quad A=J+\omega \,G,\quad
B=K-2\omega \det F.
\end{equation}
For $\gamma=2$, there exists a supplemental first integral
\begin{equation}\label{int_g2.1}
G=\sum_{i,k} F_{ik}^{2}=2(\tilde E+\omega A) t^{2}+k_1t+k_0,
\end{equation}
 where $k_0$ and $k_1$ are
constants.
%\begin{equation}\label{6}
%G=M\sinh(\sqrt{2 \omega \cos \theta}t+\phi_{0})-\frac{4(\tilde
%E+A\sin\theta)\omega}{4\omega^{2}\cos^2 \theta},\end{equation}
%\tilde C_{1}\cos{lt}+\tilde C_{2}\sin{lt}+\frac{4E+2lA}{l^{2}}
%and \eqref{int_g2} for $s=1$.
\end{theorem}

\proof To prove \eqref{5.0}
 we multiply every equation of system  \eqref{4} by  $\dot F_{ik}$
and add the results. We  get
$$(\ddot F_{ik}\dot F+\tfrac{\partial
U}{\partial F}+L\dot F+\omega^{2}\mathcal D F) \dot F=\tilde E'=0.$$

To prove the existence of integrals  $A$ and $B$ from \eqref{5} with
$k=1$ we check that
$$-F_{11}\frac{\partial U}{\partial F_{21}}-F_{12}\frac{\partial
U}{\partial F_{22}}+F_{21}\frac{\partial U}{\partial
F_{11}}+F_{22}\frac{\partial U}{\partial F_{12}}=0.$$ Indeed, we
denote $\eta={\rm det}\,F$ and get  $$-F_{11}\frac{\partial
U}{\partial F_{21}}-F_{12}\frac{\partial U}{\partial
F_{22}}+F_{21}\frac{\partial U}{\partial
F_{11}}+F_{22}\frac{\partial U}{\partial
F_{12}}=$$$$(F_{11}F_{12}-F_{12}F_{11}+F_{21}F_{22}-F_{22}F_{21})U'_{\eta}=0.$$

Then we differentiate $J$ and take into account  \eqref{4} to obtain
$$
\dot J =F_{11}\ddot F_{21}+F_{12}\ddot F_{22}-\ddot
F_{11}F_{21}-\ddot F_{12}F_{22}=\phantom{F_{11}(-l\dot
F_{11}-\frac{\partial U}{\partial F_{21}}}
$$$$
F_{11}(-l\dot F_{11}-\frac{\partial U}{\partial
F_{21}}-\omega^{2}F_{21})+F_{12}(-l\dot F_{12}-\frac{\partial
U}{\partial F_{22}}-\omega^{2}F_{22})-$$$$F_{21}\,(l\dot
F_{21}-\frac{\partial U}{\partial
F_{11}}-\omega^{2}F_{11})-F_{22}\,(l\dot F_{22}-\frac{\partial
U}{\partial F_{22}}-\omega^{2}F_{12})= -l\sum\limits_{i,j}
F_{ij}\dot F_{ij}=-\frac{l}{2}\dot G.$$ It implies
$J+\frac{l}{2}G=\mbox{const}$ and \eqref{5},  if we take into
account that $l=2\omega$.

 The proof of existence
of the integral  $B$ is analogous, since
\begin{eqnarray}-F_{11}\frac{\partial U}{\partial
F_{12}}-F_{21}\frac{\partial U}{\partial
F_{22}}+F_{12}\frac{\partial U}{\partial
F_{11}}+F_{22}\frac{\partial U}{\partial F_{21}}=0.\end{eqnarray}

To prove \eqref{int_g2.1} we notice that for $\gamma=2$ by the Euler
theorem  $\sum \limits_{i,j} F_{ij}\frac{\partial U}{\partial
F_{ij}}=2U$, since $U=U_{0}(\det F)^{-1}$ is a homogeneous function
of order $-2$.

Thus,
$$\frac{\ddot G}{2}=\sum\limits_{i,j}(\dot F_{ij}^{2} +F_{ij}\ddot
 F_{ij})=$$
 $$
=\sum\limits_{i,j} \dot F_{ij}^{2}+F_{11}\,(l\dot
F_{21}-\frac{\partial U} {\partial
F_{11}}+\omega^{2}F_{11})+F_{12}\,(l\dot F_{22}-\frac{\partial U}
{\partial F_{12}}+\omega^{2}F_{12})+$$$$F_{21}(-l\dot
F_{11}-\\\frac{\partial U} {\partial
F_{21}}+\omega^{2}F_{21})+F_{22}(-l\dot F_{12}-\frac{\partial U}
{\partial F_{22}}+\omega^{2}F_{22})=$$$$(2E-2U)+l(F_{11}\dot
F_{21}+F_{12}\dot F_{22}- F_{21}\dot F_{11}-F_{22}\dot
F_{12})-\sum\limits_{i,j}F_{ij}\frac {\partial U}{\partial
F_{ij}}+\omega^{2}G=$$$$2E+lJ-w^{2}G=2\tilde
E+lJ+2\omega^{2}G=2(\tilde E+\omega A)=\rm const.$$ It implies
\eqref{int_g2.1}. $\square$

\medskip
\begin{remark} We can see that \eqref{int_g2.1} and  \eqref{int_g2}
are similar, they are polynomials of order 2 with respect to $t$.
For $\delta=0$ the situation is absolutely different. Namely, as is
shown in \cite{T},
$$G=M\sin ( lt+\phi_{0})+\frac{4\tilde E+2lA}{l^{2}},$$
with constant $M$ and $\phi_0$, it is periodic. One can see that the
in the case $\delta=1$, $k=1$ the influence of rotation is
neutralized.
\end{remark}

\subsection{Representation of solution}

\begin{theorem} For $\delta=0$ and $k=1$ the system \eqref{4} can be
reduced to one ODE
\begin{equation}\label{8}
u'(t)=\pm \frac{1}{s^2(t)}\sqrt{f(u)}
\end{equation}
and integrated.
 Here $f(u)=D-
\tfrac{4U_{0}}{\sin{2u}}-\tfrac{A^{2}+B^{2}+2AB
\sin{2u}}{\cos^{2}{2u}},$   $D=\frac{4(\tilde E+\omega
A)k_0-k_1^2}{4}=\rm const$, $s^2(t)=G(t)$.
\end{theorem}

\proof The change of variables  $$\begin{pmatrix}
F_{11} & F_{12} \\
F_{21} & F_{22}
\end{pmatrix}=
\begin{pmatrix}
\cos v & -\sin v \\
\sin v & \cos v
\end{pmatrix}
\begin{pmatrix}
s \cos u & 0 \\
0 & s \sin u
\end{pmatrix}
\begin{pmatrix}
\cos{w} & -\sin{w} \\
\sin{w} & \cos{w}
\end{pmatrix}% \eqno(7)
$$
reduces the system of first integrals \eqref{5} to
\begin{equation}\label{9}
 \tfrac{s^{2}}{2} ((u')^{2}+\omega^{2}+(v')^{2}+2\sin(2u)v'
w'+(w')^{2})+\tfrac{(s')^{2}}{2}+\tfrac{2U_{0}}{s^{2}
\sin{2u}}=\tilde E,
\end{equation}
\begin{equation}\label{10}
(v'+\tfrac{l}{2})+\sin(2u)w'=\tfrac{A}{s^{2}},\qquad\qquad
\sin(2u)(v'+\tfrac{l}{2})+w'=-\tfrac{B}{s^{2}}.
\end{equation}
Further, $G=\sum_{i,k} F_{ik}^{2}=s^{2}(t)$, therefore we can use
the expression \eqref{int_g2.1}.

From \eqref{10} we obtain
$$v'=\frac{1}{s^{2}}\cdot\frac{A+B\sin{2u}}{\cos^{2}{2u}}-\frac{l}{2},$$
$$w'=-\frac{1}{s^{2}}\cdot\frac{B+A\sin{2u}}{\cos^{2}{2u}}.$$
Together with  \eqref{9} it implies \eqref{8}. $\square$

\medskip

\begin{remark} The solution $u(t)$ of \eqref{8} contains elliptic
integrals. Indeed, if we denote $r=\sin{2u}$, then $\dot
u=\frac{\dot r}{2\sqrt{1-r^{2}}}$ and $\cos^{2}{2u}=1-r^{2}$. Thus,
\eqref{8} can be reduced to $\frac{\dot
r}{2}=\pm\frac{1}{s^{2}}\sqrt{\frac{f(r)}{r}},$ where $f(r)$ is a
polynomial of the third order. Thus,
$$\int\limits_0^t \frac{1}{s^2(\tau)} d\tau=$$
$$=\mp\tfrac{r_{1}}{\sqrt{-Dr_{2}(r_{3}-r_{1})}}\left(F(\sqrt{\tfrac{r(r_{3}-r_{1})}
{r_{3}(r-r_{1})}},\sqrt{\tfrac{r_{3}(r_{1}-r_{2})}
{r_{2}(r_{1}-r_{3})}})-\Pi(\tfrac{r_{3}}{r_{1}-r_{3}},
\sqrt{\tfrac{r(r_{3}-r_{1})} {r_{3}(r-r_{1})}},
\sqrt{\tfrac{r_{3}(r_{1}-r_{2})} {r_{2}(r_{1}-r_{3})}})\right),$$
where $r_{1}$, $r_{2}$, $r_{3}$ are the roots of equation $f(r)=0$,
$F$ and $\Pi$ are normal elliptic Legendre integrals of the first
and third order in the Jacobi form. The roots are simple, otherwise
among them there is one that corresponds to equilibrium and the
solution does not exist in its neighborhood.
\end{remark}

\begin{remark}
It is easy to check that if we set
 $s^2=G=2Et^{2}+ k_1 t+k_0$ and $l=\omega=0$, then
$D=\tfrac{8Ek_0-k_1^{2}}{4}$. Thus, we get the result of of
\cite{AnisimovLysikov}, where the left-hand side of the formula is
$\int\tfrac{dt}{2Et^{2}+k_1
t+k_0}=\tfrac{1}{2E\beta}\arctan\tfrac{4Et+k_1}{\beta}$, with
$\beta^{2}=4D>0$.
\end{remark}

\begin{remark}
If we know the solution to \eqref{8}, we can go back to the Eulerian
coordinates and find  ${\bf u}$. Indeed, the functions $v$ and $w$
can be found from \eqref{10}. Thus, the coefficients of the matrix
$F$ are known,  $Q=\dot F F^{-1}$ and ${\bf u}=Q{\bf x}$.
\end{remark}

\begin{remark} If $u$ tends to zero, then $\det F $ tends to zero and the components of $Q$ tends to infinity.
Since these components have sense of derivatives of the solution,
then it implies a blow-up.  Thus, for the model of vortex with
uniform deformation the instability of an equilibrium in fact
implies blow-up of derivatives. One can hypothesize that the same
phenomenon holds for a localized vortex in a gas, i.e. the
instability leads to formation of singularities (see \cite{RT2}).
\end{remark}

\section {Discussion}

We prove that if we take into account a small correction due to
centrifugal force usually neglected in the  $l$-plane model of
atmosphere, we drastically change properties of a specific class of
vortices. However, this does not mean that the generally accepted
model is erroneous. It is well known that it is approximate and
adequately describes only processes of relatively small scale. The
model we are considering is also approximate, and an accurate
description of the processes of the vortex dynamics of the scale of
tropical cyclones can be carried out only within the framework of
equations on a sphere.

In addition, vortex solutions with a uniform deformation, on which
the difference in stability is manifested, have infinite energy.
That is, they are nonphysical on the whole plane. To notice the
effect described in this paper, it is necessary that the members
${\mathcal L}{\bf u}$ and $\omega^2 {\bf x})$ (see \eqref{(1)}) have
the same order. For this, the linear velocity profile must be
maintained inside the vortex within a radius of the order of $
\frac{l}{\omega^2} {\bf u}$ $ \approx 100$ km (the typical value of
velocity ${\bf u}$ is about $10$ m/sec). This is not observed in
nature, in fact, this radius is about 30 km. Near the equator, where
the Coriolis parameter is small, stable atmospheric vortices are not
observed. Perhaps the effect described in this paper may be one of
the factors of their destabilization.

Solutions with a linear velocity profile may seem only an
interesting mathematical object. However, it is not quite like that.
In fact, solutions of this class have enormous applications in
astrophysics (e.g.\cite{Bogoyavlensky}), in the point blast theory
 \cite{Korobeinikov}, etc. For us, their
application to geophysics is important. The motivation is the fact
that, near the center of a large atmospheric vortex in its
conservative stage, the structure of wind velocity is linear. There
are attempts to use this fact to predict the motion of the
atmospheric vortex \cite{RYH2010}, \cite{RYH2012}. Note that the
full structure of stationary vortices is very diverse \cite{R15}.

\section*{Acknowlegments}
The first author thanks Nikolai Leontiev for a stimulating
discussion.

\end{document}